\documentstyle[epsfig]{article}
\begin{document}
\def\b{\bar}
\def\d{\partial}
\def\D{\Delta}
\def\cD{{\cal D}}
\def\cK{{\cal K}}
\def\f{\varphi}
\def\g{\gamma}
\def\G{\Gamma}
\def\l{\lambda}
\def\L{\Lambda}
\def\M{{\Cal M}}
\def\m{\mu}
\def\n{\nu}
\def\p{\psi}
\def\q{\b q}
\def\r{\rho}
\def\t{\tau}
\def\x{\phi}
\def\X{\~\xi}
\def\~{\widetilde}
\def\h{\eta}
\def\bZ{\bar Z}
\def\cY{\bar Y}
\def\bY3{\bar Y_{,3}}
\def\Y3{Y_{,3}}
\def\z{\zeta}
\def\Z{{\b\zeta}}
\def\Y{{\bar Y}}
\def\cZ{{\bar Z}}
\def\`{\dot}
\def\be{\begin{equation}}
\def\ee{\end{equation}}
\def\bea{\begin{eqnarray}}
\def\eea{\end{eqnarray}}
\def\half{\frac{1}{2}}
\def\fn{\footnote}
\def\bh{black hole \ }
\def\cL{{\cal L}}
\def\cH{{\cal H}}
\def\cF{{\cal F}}
\def\cP{{\cal P}}
\def\cM{{\cal M}}
\def\olam{\stackrel{\circ}{\lambda}}
\def\oX{\stackrel{\circ}{X}}
\def\oX{\stackrel{\circ}{X}}
\def\const{{\rm const.\ }}
\def\ik{ik}
\def\mn{{\mu\nu}}
\def\a{\alpha}
\input{axodraw.sty}

\title{Kerr geometry beyond the Quantum theory}

\author{A. BURINSKII \\
\\
Gravity Research Group, NSI Russian Academy of Sciences,\\
B. Tulskaya 52, 115191 Moscow, Russia}

\date{\it Invited talk at the Int. conference ``Beyond the Quantum'',
Leiden, May 2006}

\maketitle

\begin{abstract}
 Dirac electron theory and QED do not take into account
gravitational field, while the corresponding Kerr-Newman solution
with parameters of electron has very strong stringy, topological
and non-local action on the Compton distances, polarizing
space-time and deforming the Coulomb field.  We discuss the relation
of the electron to the Kerr's microgeon model and argue that the
Kerr geometry may be hidden beyond the Quantum Theory. In
particular, we show that the Foldi-Wouthuysen `mean-position'
operator of the Dirac electron is related to a complex
representation of the Kerr geometry, and to a complex stringy
source. Therefore, the complex Kerr geometry may be hidden beyond
the Dirac equation.

\end{abstract}

\section{Introduction}\label{aba:sec1}

Superstring theory \cite{GSW} is based on the extended stringy
elementary states: $\quad  Points  \quad \longrightarrow \quad
Extended \ Strings,$ and also, on the unification of the Quantum
Theory with Gravity on Planckian level of masses $ M_{pl},$ which
correspond to the distances of order $10^{-33}$ cm. Such a
penetrating into the deep structure of space-time has been based
on the convincing evidences:

a/ The brilliant confirmation of the predictions of QED which
ignored  gravitational field and has been tested up to the distances
of order $10^{-16}$ cm. It suggests that boundary of Quantum Gravity
may be shifted at least beyond the distances $10^{-16}$ cm.

b/ The dimensional analysis, showing that $M_{pl}$ corresponds to
the energies $E_{pl} = \sqrt{\hbar c^5/G}$ which are formed from
the fundamental constants relating quantum theory, $\hbar,$
special relativity, $c,$ and gravity, $G$.

c/ Estimation of the masses $M_q$ and distances, where the action
of gravity may be comparable with the action of quantum effects,
which is done by the comparison of the gravitational
radius of the Schwarzschild black hole $r_g =2M_q$  with the
Compton radius of the corresponding quantum particle $r_c = 1/M_q$
(we use here the Planck units $\hbar=c=G=1$). One sees that the
equality $r_g\sim r_c$ is achieved by the Planckian masses $M_q
\sim 1 ,$ i.e. by $M_q \sim 10^{-33} cm .$ It leads to the
conclusion that quantum gravity has to act on the Planckian scale
$r_g \approx r_{c} = \frac 1 {M_q} \sim 1.$

All that is convincing, except for the argument c/ . The
Schwarzschild geometry does not take into account  spin of quantum
particles which is indeed very high with respect to the masses. In
particular, for electron $S= 1/2 ,$ while $ m\approx 10^{-22}.$
So, to estimate gravitational field of spinning particle, one has
to use the Kerr, or Kerr-Newman solutions \cite{DKS}. Of course,
there may be objections that quantum processes are strongly
non-stationary because of the vacuum
fluctuations, and they cannot
be described by the stationary Kerr and Kerr-Newman solutions.
However, QED tells us that electromagnetic radiative corrections
are not too large. On the other hand, we do not know another
solution which could better describe the gravitational field of a
spinning particle. In any case, estimations on the base Kerr
solution have to be much more correct then on the base of
Schwarzschild solution.

{\it Performing such estimation, we obtain a striking
contradiction with the above scale of Quantum Gravity !}

\smallskip

Indeed, for the Kerr and Kerr-Newman solutions we have the basic
relation between angular momentum $J$, mass $m$ and radius of the
Kerr singular ring $a$ : \be J=ma \label{Sma}. \ee Therefore,
Kerr's gravitational field of  a spinning particle is extended
together with the Kerr singular ring up to the distances $a= J/m
=\hbar /2m \sim  10^{22}$ which are of the order of the Compton
length of electron $10^{-11}$ cm., forming a singular closed
string \fn{See also
\cite{Isr,Bur0,Lop,BurOri}.} Therefore, in analogy with
string theory {\it the `point-like' Schwarzschild singularity
turns in the Kerr geometry into an extended  string of the
Compton size. }

Notice, that the Kerr string is not only analogy. It was shown
that the Kerr singular ring is indeed the string \cite{BurOri},
and, in the analog of the Kerr solution to low energy string theory
\cite{Sen}, the field around the Kerr string is similar to the
field around a heterotic string \cite{BurSen}.

The use of Kerr geometry for estimation of the scale of Quantum
Gravity gives the striking discrepancy with respect to the
estimation done with the Schwarzschild solution. We arrive at the
conclusion that the Kerr geometry has to play an important role in
Quantum processes on the Compton distances of electron, of order
$\sim 10^{-11}$ cm. The local gravitational field at these
distances is extremely small, and  the strong field is
concentrated near an extremely narrow vicinity of the Kerr
singular ring which forms a closed
 string of the Compton radius.

\section{Real structure of the Kerr-Newman solution.}

The Kerr-Newman metric may be represented in the Kerr-Schild form
\be g_\mn =\eta_\mn -2H k_\m k_\n \label{KS}, \ee where $\eta
_\mn$ is auxiliary Minkowski metric and $H=\frac {mr - q^2/2}
{r^2+ a^2 \cos^2\theta} .$

One sees that metric is Minkowskian almost everywhere, for
exclusion of the negligibly small subset of the space-time.
However, this stringy subset has very strong dragging effect which
polarizes space-time leading to a very specific polarization of
the electromagnetic fields. As a result, the electromagnetic field
of  the corresponding Kerr-Newman solution $F_\mn$, which cannot
be consider as a weak one for parameters of charged particles,
turns out to be aligned with the Kerr principal null congruence.
Electromagnetic and gravitational fields are formed by the
twisting vector field $k_\m (x),$ principal null congruence (PNC),
and acquire the Kerr stringy circular singularity as a caustic of
PNC.

The explicit form of the field $k_\m$ is determined by the
one-form
 \be k_\m dx^\m = dt +\frac z r dz + \frac r {r^2 +a^2} (xdx+ydy) - \frac a
{r^2 +a^2} (xdy-ydx) .  \label{km} \ee It is a twisting family of
null rays, fig.1, forming a vortex which is described by the Kerr
theorem in twistor terms\cite{BurNst,Multi,KraSte}.\fn{Complicate form of
the field $k_\m(x)$ determines the complicate form of the Kerr
metric, contrary to the extremely simple Kerr-Schild
representation (\ref{KS}).}

\begin{figure}[t]
\begin{center}
\psfig{file=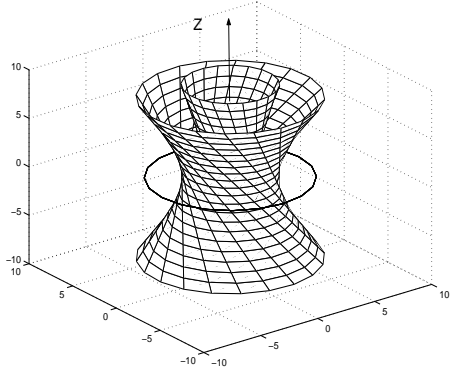,height=5cm,width=6cm}
\end{center}
\caption{The Kerr ring as a branch line, and the twistor lines of
the Kerr congruence which covers twice the space-time.}
\end{figure}

PNC plays very important role, since the field $k_\m$ determines
not only the form of Kerr-Newman metric with mass $m$ and charge
$q, $ but also the Kerr-Newman electromagnetic vector potential $
A_\m =\frac {q r} {r^2+ a^2 \cos^2 \theta}  \ k_\m, \label{Am} ,$
and the flow of radiation in the radiative r
otating solutions\fn{

$\Phi(r, \phi, \theta)=\frac 1 {r^2 + a^2 \cos ^2 \theta} [-6
m(\ddot x_0^\m k_\m) + 2 \dot m]$ is an angular distribution of
the energy density of radiation. Radiation may be related to the
loss of mass $\dot m <0$, acceleration $\ddot x_0^\m k_\m \ne 0$
and to the wave electromagnetic excitations of the  Kerr-Newman
solution \cite{BurNst,BurOri,KraSte}.} $ T_\mn = \Phi (r, \phi, \theta)
k_\m k_\n \label{Rad}. $

The congruence covers spacetime twice, and the Kerr ring is a
branch line of the space on two sheets: positive sheet of the
`outgoing' fields ($r>0$) and negative sheet of the `ingoing'
fields. ($r<0$). Notice, that for $a^2>>m^2$ the black hole
horizons are absent, and space-time acquires a twofold topology
\cite{Isr,Lop}.

\smallskip

{\it There appears the Question: ``Why  Quantum Theory does not
feel such drastic changes in the structure of space time on the
Compton distances?''}

\smallskip

How can such drastic changes in the structure of  space-time and
electromagnetic field  be experimentally unobservable and
theoretically ignorable in QED?

The negative sheet of Kerr geometry may be truncated along the
disk $r=0$. In this case, inserting the truncated space-time into
the Einstein-Maxwell equation, one obtains on the `right' side of
the equations the source  with a disk-like support. This source
has a specific matter with superconducting properties
\cite{Isr,Lop}. The `negative' sheet of space appears now as a
mirror image of the positive one, so the Kerr  singular ring is an
`Alice' string related to the mirror world. Such a modification
changes interpretation, but does not simplify problem, since it
means that Quantum Theory does not feel this `giant' mirror of the
Compton size, while the virtual charges have to be very sensitive
to it.\fn{Note, that this disk is relativistically rotating and
has a thickness of the order of classical size of electron,
$r_e=e^2/2m ,$\cite{Lop,renorm}.}

The assumption, that QED has to be corrected taking into account
the peculiarities of
 the space-time caused by the Kerr geometry,
may not be considered as reasonable because of the extraordinary
exactness of the QED.

There is apparently the unique way to resolve this contradiction:
to conjecture that  the Kerr geometry is hidden beyond the Quantum
Theory, i.e. is already taken into account and play there
essential role.

From this point of view there is no need to quantize gravity,
since the Kerr geometry may be the source of some quantum
properties, i.e. may be primary with respect to the Quantum
Theory.

\section{Microgeon with spin}

Let us consider the Wheeler's model of {\it Mass Without Mass --
`Geon'.} The photons are moving along the ring-like orbits, being
bound by the own gravitational field. Such field configuration may
generate the particle-like object having the mass and angular
momentum. Could such construction be realized with an unique
photon? In general, of course - not, because of the weakness of
gravitational field. However, similar idea on `mass without mass'
is realized in the theory of massless relativistic strings and may
be realized due to the stringy properties of the Kerr solution
with $a>>m .$ In the Kerr geometry, one can excite the Kerr
circular string by an electromagnetic field  propagating along
this singular string as along of a waveguide. Electromagnetic
excitations of the Kerr source with $a>>m$ has the stringy
structure, and  leads to a contribution to the mass and spin. In
particular, the model of microgeon with spin turns out to be
self-consistent \cite{Bur0,BurOri,BurTwi}.

Analysis of the exact {\it aligned} electromagnetic excitations on
the Kerr background shows an unexpected peculiarity:
{\it the inevitable appearance of two axial
singular semi-infinite half-strings of opposite chiralities}.
There appears the following stringy skeleton of a spinning particle,
fig.  2.

\begin{figure}[t]
\begin{center}
\psfig{file=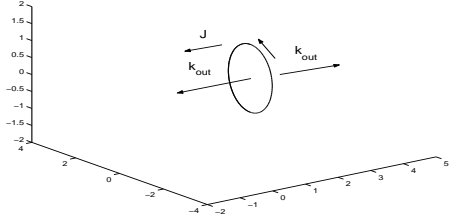,height=3.5cm,width=6cm}
\end{center}
\caption{Skeleton of the Kerr Spinning Particle.}
\end{figure}

The spin of this microgeon may
 be interpreted as excitation of the
Kerr string by a photon moving along a circular orbit, which is
reminiscent of the electron self-energy diagram  in QED
\medskip

{\small
\begin{picture}(150,100)(-50,0)
\ArrowLine(45,70)(0,70) {} \ArrowLine(115,70)(75,70) {}
\DashArrowArcn(60,70)(15,180,0)3 {}
\LongArrowArcn(60,70)(15,0,-90) {}
\LongArrowArcn(60,70)(15,-90,-180) {}
%\DashArrowLine(60,15)(60,55)3 {}
\end{picture}}.

In the Kerr's gravity, the virtual photon line of this diagram
does not leave the Compton region of the particle due to the Kerr
stringy waveguide. As it was shown in \cite{BurTwi}, the axial
half-strings are the null-strings (the Schild, or the pp-wave
strings) and may be described by the bilinear spinor combinations
formed from the solutions of the Dirac equation.\fn{The axial and
circular singularities form a specific multischeeted topology of
space-time, which admits the spinor two-valuedness.}

Moreover, there is a wonderful fact, that the basic quantum
relation $E=\hbar \omega$ is already contained in the basic
relation of the Kerr geometry $J=ma$, (\ref{Sma}). Indeed, setting
$J=\frac \hbar 2 ,$  one writes  (\ref{Sma}) as
 $ a=\frac \hbar {2m} . $

So far, we considered the constant $\hbar$ not as a quantum
constant, but as an experimentally constant characterizing the
spin of electron. Let us consider now the classical fields
propagating along the Kerr ring with speed of the light and with
the winding number of phase $n=1/2.$ The corresponding length of
wave will be $\lambda =2\pi a / n = 2\pi \hbar/m $ and the
corresponding frequency $\omega = 2\pi c/\lambda = c m /\hbar .$
It yields \be \frac E c\equiv m c = \hbar \omega  . \label{Q}\ee
 Up to now, we
have not used the quantum operators at all. We have used only the
topological properties providing the two-valued representations by
rotations, or the classical quantization of phase (winding
number). As a result, we have obtained the quantum relation
(\ref{Q}) from the classical Kerr relation $J=ma.$ It suggests
that {\it `Kerr's geometry'
 may cause the origin of Quantum
properties,}

\section{Dirac equation and the complex Kerr geometry}

\subsection{Dirac equation in the Weyl basis}

 In the Weyl basis the Dirac equation demonstrates some interesting
 peculiarities.\fn{We use spinor notations of the book
 \cite{WesBag}, see also\cite{BLP}.
 The Dirac matrices in the Weyl basis take the form

$ \gamma^\m  = \left( \begin{array}{cc}
0 & \sigma ^\m \\
\bar\sigma ^\m & 0
\end{array} \right) \ ,$ where
$\bar\sigma ^{\m \dot\alpha \alpha} =
\epsilon ^{\dot \alpha \dot \beta} \epsilon ^{ \alpha  \beta}
\sigma ^\m _{\beta \dot\beta},$ and
$\sigma _0  =\bar\sigma _0, \quad \sigma _k  = - \bar\sigma _k , \ k=1,2,3 ,$
\begin{equation}
\sigma _0=
\left( \begin{array}{cc}
1 & 0 \\
0 & 1
\end{array} \right)  , \quad
\sigma _1  =
\left( \begin{array}{cc}
0 & 1 \\
1 & 0
\end{array} \right)  , \quad
\sigma _2  =
\left( \begin{array}{cc}
0 & -i \\
i & 0
\end{array} \right) , \quad
\sigma _3  =
\left( \begin{array}{cc}
1 & 0 \\
0 & -1
\end{array} \right) .
\end{equation}}
The Dirac spinor has the form
 $\Psi =
\left(\begin{array}{c}
 \phi _\alpha \\
\chi ^{\dot \alpha}
\end{array} \right),$
and the Dirac equation splits into
\be\sigma ^\m _{\alpha \dot \alpha} (i \d_\m  +e A_\m)
 \chi ^{\dot \alpha}=  m \phi _\alpha , \quad
 \bar\sigma ^{\m \dot\alpha \alpha} (i \d_\m  +e A_\m)
 \phi _{\alpha} =  m \chi ^{\dot \alpha}.\label{Wspl} \ee
 Conjugate spinor has the form
 \be \bar\Psi =(\chi ^+,\phi ^+ )=(\bar \chi^\alpha, \bar
 \phi_{\dot\alpha}). \ee

The Dirac current \be J_\m = e (\bar \Psi \gamma _\m \Psi) = e
(\bar\chi  \sigma _\m  \chi + \bar\phi  \bar \sigma _\m  \phi ),
\ee  can be represented as a sum of two lightlike components of
opposite chirality \be J^\m_{\mathrm{L}} = e \bar\chi \sigma^\m \chi \ ,
\qquad J^\m_{\mathrm{R}} = e \bar\phi \bar\sigma^\m \phi. \ee

The corresponding null vectors $ k^\m_{\mathrm{L}} = \bar\chi \sigma^\m \chi
\ ,$ and $k^\m_{\mathrm{R}} = \bar\phi \bar\sigma^\m \phi $, determine the
considered above directions of the lightlike
half-strings in the
microgeon model. The momentum of the Dirac electron is $p^\m =
\frac m 2 (k^\m_{\mathrm{L}} + k^\m_{\mathrm{R}}),$
and the vector of polarization of
electron \cite{AkhBer,BLP} in the state with a definite projection
of spin on the axis of polarization  is
$n^\m = \frac 12 (k^\m_{\mathrm{L}} -
k^\m_{\mathrm{R}}).$ In particular, in the rest frame and the axial
z-symmetry, we have $k_{\mathrm{L}}=(1,\vec
k_{\mathrm{L}})=(1,0,0,1)$ and $k_{\mathrm{R}}=(1,\vec
k_{\mathrm{R}})=(1,0,0,-1),$ which gives
 $p^\m = m(1,0,0,0),$ and $n^\m =  (0,0,0,1),$
which corresponds to the so-called transverse polarization of
electron \cite{AkhBer}, $\vec n \vec p =0 .$

The Dirac spinors form a natural null tetrad. The null vectors $
k^\m_{\mathrm{L}} = \bar\chi \sigma^\m \chi $ and $
k^\m_{\mathrm{R}} = \bar\phi \bar\sigma^\m \phi $, may be
completed to the null tetrad by two null vectors $ m^\m = \phi
\sigma^\m \chi \ ,$ and $\bar m^\m = (\phi \sigma^\m \chi)^+ $
which are controlled by the phase of wave function. Therefore, the
de Broglie wave sets a synchronization of the null tetrad in the
surrounding space-time, playing the role of an `order parameter'.

It is well known \cite{Car} that the Kerr-Newman solution has the
same gyromagnetic ratio ($g=2$), as that of the Dirac electron.
There appears a natural question: is it an accidental coincidence,
or there is a deep relationship between the Dirac equation and the
Kerr-Newman geometry? This problem is related to the problem of
description of electron in coordinate representation, and to the
problem of localized states in the Dirac theory
\cite{NewWig,Schw,FolWou}.

\subsection{The problem of position operator in the Dirac theory}

It is known \cite{Schw} that the position operator  $\hat {\vec
x}= \nabla _{\vec p} $ is not Hermitean in any relativistic
theory, $(\Psi,\hat x \Phi) \ne (\hat x \Psi,\Phi).$ In the Dirac
theory, the  problem of the operator coordinate is still more
complicate.
The plane wave solutions of the Dirac equation correspond to the
state
s with a fixed momentum, and the position of electron
$\vec x$ is undetermined for the plane waves.\fn{To have a localization of position, one needs to form a wave
packet.}

 The velocity $\dot x$  for the operator of coordinate $x$ is $\dot
x =(xH-Hx)= c\alpha ,$ and projection of the velocity on any
direction yields $\pm c .$ Dirac shows \cite{Dir} that this
equation may be integrated, yielding for coordinate $x$ the result

\be x= -\frac 14 c\hbar ^2 \dot \alpha _x ^0 e^{-2i Ht/\hbar }
H^{-2} + c^2 p_x H^{-1}t + x_0 .\label{x} \ee

Therefore, the velocity of coordinate $x$ consists of  the
constant term $c^2 p_x H^{-1}$ and the oscillating contribution $
\frac 12 ic\hbar \dot \alpha _x ^0 e^{-2i Ht/\hbar } H^{-1} , $
the so-called `zitterbewegung'.

In the rest frame, $\vec p=0 ,$ one can get $i\hbar \dot \alpha
_x= -2 H\alpha _x ,$ and the oscillating part $ \tilde x= -\frac
14 c\hbar ^2 \dot \alpha _x ^0 e^{-2i H t/\hbar } H^{-2} $
 has the Compton amplitude $\frac 12 ic\hbar \alpha _x
H^{-1}.$

Since this expression describes complex oscillations, it is more
convenient to consider the complex combination \be \tilde x+i
\tilde y =\frac 14 ic\hbar ^2 (\dot\alpha _x^0 +i \dot\alpha
_y^0)e^{-2i H t/\hbar } H^{-2} \label{xy} \ee which describes
circular motion in the $(xy)$-plane along a ring of the Compton
size $ \frac {c\hbar}{2m}.$ These oscillations take place in the
plane which is orthogonal to the polarization direction of the
electron. One can see here a correspondence to the structure of
the Kerr microgeon model.

The `zitterbewegung' problem is related to the problem of
localized states and leads to the treatment of a
`mean-position' operator \cite{NewWig}.
 The best solution in this problem was found by Foldi and
 Wouthuysen\cite{FolWou,Schw} which performed a very nontrivial
 matrix transformation of the wave function $\Psi$ and Hamiltonian
 $H$ to the so-called  Foldi-Wouthuysen (FW) representation,
  $\Psi _{\mathrm{FW}} = e^{iS_{\mathrm{FW}}} \Psi $ and $H_{\mathrm{FW}} = e^{iS_{\mathrm{FW}}}H e^{
-iS_{\mathrm{FW}}} .$

   In the new representation the
  negative frequency modes are suppressed, `zitterbewegung' is absent,
  and the new (rather complicate) position operator
  $\vec X_{\mathrm{FW}}=e^{iS_{\mathrm{FW}}}\vec x e^{-iS_{\mathrm{FW}}}$ corresponds to the
  conventional concept of the velocity of particle,
  $\dot {\vec X}_{\mathrm{FW}}=i[H_{\mathrm{FW}},\vec x]= \beta \vec p /E_{\vec p} .$

On the other hand, the use of ordinary position operator $\vec x $
in the Foldi-Wouthuysen representation may be transformed back to
the Dirac or Weyl representation $\hat {\vec X} =
e^{-iS_{\mathrm{FW}}} \vec x e^{iS_{\mathrm{FW}}}$, leading to a
complicate mean-position operator without `zitterbewegung'. This
operator  is simplified in the rest frame of the electron, $\vec p
=0$, and takes the form \be \hat X^\m = x^\m + i  \frac {\hbar c}
{2m} \gamma ^\m  \label{hatX}.\ee The resulting coordinate of
electron turns out to be complex. In the terms of the null vectors
$k_{\mathrm{L}}=(1,\vec k_{\mathrm{L}})=(1,0,0,1)$ and
$k_{\mathrm{R}}=(1,\vec k_{\mathrm{R}})=(1,0,0,-1),$ it takes the
form \be (\bar\Psi\hat X \Psi) =x + i a
(k_{\mathrm{L}}+k_{\mathrm{R}}) \label{X} , \ee where x is a
center of mass and $a= \frac {\hbar c}{2m}$ is the Compton length.
In the Weil representation, the vectors $k_{\mathrm{L}}$ and
$k_{\mathrm{R}}$ transform independently by Lorentz
transformations and transfer to each other by the space reflection
(inversion) $P= \eta_P \gamma_ 4, \ |\eta _P|=1.$ It gives a hint
that the Dirac particle may be formed by two complex point-like
particles $X_{\mathrm{L}}$ and $X_{\mathrm{R}}$ propagating along
the complex world-lines \be X^\m_{\mathrm{L}}(t)=x^\m (t) + ia
(1,0,0,1),  \quad X^\m_{\mathrm{R}}(t)=x^\m (t) + ia (1,0,0,-1) ,
\label{XDLR} \ee where the 3-vector of mean-position is $\vec
X=\frac 12 (\vec X_{\mathrm{L}}+\vec X_{\mathrm{R}})=\vec x(t) .$

Such a representation turns out to be close related to the
known complex representation of the Kerr geometry
\cite{New,Bur0,BurStr,BurNst,B
urTwi}.

\subsection{Complex representation of the Kerr geometry}

In 1887 (!) Appel \cite{appel} consider   a simple complex
transformation of the Coulomb solution $\phi= q/r, $  a complex
shift $(x,y,z) \to (x,y,z+ia)$ of the origin
$(x_0,y_0,z_0)=(0,0,0)$ to the point $(0,0,ia).$ On the real
section (real$(x,y,z)$), the resulting solution
\be \phi(x,y,z)=
\Re e \ q/\tilde r \ee

acquires a complex radial coordinate
$\tilde r =\sqrt{x^2+y^2+(z-ia)^2}.$ Representing $\tilde r$ in
the form \be \tilde r= r-ia \cos \theta \label{tra} \ee
 one obtains
for $\tilde r ^2$

\be r^2-a^2\cos ^2\theta - 2ia r \cos \theta = x^2+y^2 +z^2 -a^2
-2iaz.\label{tr2}\ee

Imaginary part of this equation gives $ z=r\cos \theta ,$ which
may be substituted back in the real part of (\ref{tr2}). It leads
to the equation $x^2+y^2 =(r^2 +a^2)\sin ^2 \theta,$ which may be
split into two conjugate equations $x\pm iy = (r \pm ia) e^{\pm
i\phi} \sin \theta.$ Therefore, we obtain the transfer from the
complex coordinate $\tilde r$ to the Kerr-Schild coordinate system
\bea x+iy &=& (r + ia) e^{i\phi} \sin \theta ,\label{KSc}  \\
\nonumber z&=&r\cos \theta , \\ \nonumber t&=& r +\rho. \eea Here
$r$ and $\theta$ are the oblate spheroidal coordinates, and the
last relation is a definition of the real retarded-time coordinate
$\rho$. The Kerr-Schild coordinates $\theta,\phi,\rho$ fixe a null
ray in $M^4$ (twistor) which is parametrized by coordinate $r.$

One sees, that after complex shift, the singular point-like source
of the Coulomb solution turns into a singular ring corresponding
to $\tilde r=0,$ or $r=\cos\theta=0.$ This ring has radius $a$ and
lies in the plane $z=0.$ The space-time is foliated on the null
congruence of twistor lines, shown on fig. 1. It is twofolded,
having the ring-like singularity as the branch line. Therefore,
for the each real point $(t,x,y,z) \in {\bf M^4}$ we have two
points, one of them is lying on the positive sheet of space,
corresponding to $r>0$, and another one lies on the negative
sheet, where $r<0$.

It was obt
ained \cite{Bur0} that the Appel potential corresponds  exactly to
electromagnetic field of the Kerr-Newman solution written on the
auxiliary Minkowski space of the Kerr-Schild metric (\ref{KS}).
 The vector of complex shift $\vec a=(a_x,a_y,a_z)$
corresponds to the direction of the angular momentum $J$ of the
Kerr solution \cite{Bur0,BurMag} and $ |a| = J/m .$

Newman and Lind \cite{New} suggested a description of the
Kerr-Newman geometry  in the form of a retarded-time construction,
in which it is generated by a complex source, propagating along a
{\it complex world line} $X^\m(\t)= x^\m(0) + u^\m \t + ia^\m $ in
a complexified Minkowski space-time $\mathbf{CM}^4$. Here time is
complex, $\t=t+i\sigma ,$ and $u^\m$ is a unit time-like vector.
The rigorous description of this representation may be given in
the Kerr-Schild approach \cite{DKS} based on the Kerr theorem and
the Kerr-Schild form of metric (\ref{KS})\fn{It is related to the
existence of auxiliary Minkowski metric $\eta^\mn ,$ which is
necessary for the complex representation, as well as for the
conditions of the Kerr theorem\cite{BurNst,Multi}.}
In the rest frame one can
consider the `left' and `right' complex world lines, related to the
complex conjugate sources $i\vec a$ and $-i\vec a.$ \be
X^\m_{\mathrm{L}}(\t_{\mathrm{L}})=x^\m (\t_{\mathrm{L}}) + i\vec
a, \quad X^\m_{\mathrm{R}}(\t_{\mathrm{R}})=x^\m (\t_{\mathrm{R}}) -
i\vec a  , \label{XDLR} \ee

The complex retarded time  is determined in analogy with the real
one, but is to be based on the complex null cones, see
\cite{New,BurNst,BurTwi}.

Let's consider the complex radial distance from a real point $x$
to a complex point $X_{\mathrm{L}}$ of the `left' complex world-line \be
\tilde r _{\mathrm{L}} =\sqrt{(\vec x    - \vec X_{L})^2} = r_{\mathrm{L}}- ia \cos
\theta _{\mathrm{L}} \label{trL}.  \ee

To determine a retarded-time parameter $\t_{\mathrm{L}}$ one has to write
down the light-cone equation $ds^2=0,$ or

\be \tilde r _{\mathrm{L}} ^2 - (t-\t_{\mathrm{L}})^2 =0 \label{dsL} \ee

It may b
e split into two retarded-advanced-time equations $t-\t_{\mathrm{L}}
= \pm  \tilde r _{\mathrm{L}} .$ The retarded-time equation corresponds to
the sign $+$ and, due to (\ref{tra}), leads to relation

\be \t _{\mathrm{L}} = t - r _{\mathrm{L}} + ia \cos \theta _{\mathrm{L}} . \label{tauL} \ee One
sees that $\t _{\mathrm{L}} $ turns out to be complex \be \t _{\mathrm{L}} = \rho _{\mathrm{L}} + i
\sigma _{\mathrm{L}} , \quad \sigma _{\mathrm{L}} = a\cos \theta _{\mathrm{L}} . \label{sigma} \ee

\subsection{Complex worldline as a string}

In the complex retarded-time construction, the left complex world
line $X_{\mathrm{L}}(\t_{\mathrm{L}})$ has to be parametrized by complex parameter $\t
_{\mathrm{L}}=\rho _{\mathrm{L}} +i \sigma _{\mathrm{L}} .$ It has a few important consequences.

i/ Being parametrized by two parameters $\rho$ and $\sigma$, a
complex world-line $X(\t)$ is really a world-sheet and corresponds
to a {\it complex string.} This string is very specific, since it
is extended in the complex time direction $\sigma.$

ii/ A fixed value of $\sigma _{\mathrm{L}}$ corresponds to the fixed value of
$\cos \theta _{\mathrm{L}} ,$ and, in accordance with (\ref{KSc}), together
with the fixed parameter $\phi$, it selects a null ray of the Kerr
congruence (twistor).

iii/Since $|\cos \theta| \le 1 ,$ parameter $\sigma$ is restricted
by interval $\sigma \in [-a,a],$ i.e. complex string is open and
the points $\rho \pm ia$ are positioned at its ends. The
world-sheet represents an infinite strip:  $(t,\sigma) : -\infty
<t < \infty, \sigma \in [-a,a] .$

iv/ From (\ref{XDLR}) and (\ref{tauL}) one sees that the left
complex point of the Dirac x-coordinate
$X_{\mathrm{L}}=ia(1,0,0,1)$ has $\Im m \ \t _{\mathrm{L}} =ia
\cos \theta _{\mathrm{L}},$ which yields $\cos \theta
_{\mathrm{L}} = 1 . $

Therefore, this is the boundary point of the complex world line,
and coordinate relations (\ref{KSc}) show that the complex light
cones positioned at this boundary have the {\it real section}
along the Kerr axial half-string $z
=r,\quad x=y=0.$

Similar treatment for the right complex point of the Dirac
x-coordinate $X_{\mathrm{R}}=ia(1,0,0,-1)$ show that it is also
placed on the same boundary of the stringy strip (the same
timelike component $ia$), however, $\Im m \ \t _{\mathrm{R}} =-ia
\cos \theta ,$ which yields $\cos \theta _{\mathrm{R}} =-1$ and
corresponds to the axial half-string propagating in opposite
direction $z=-r,\quad x=y=0.$

Therefore, in the real space-time the two complex sources of the
Dirac operator of coordinate have the real image in the form of
the considered above two axial semi-infinite half-strings: left
and right.\fn{For more details see \cite{BurStr,BurOri,BurTwi}.}

Note, that there is an asymmetry in the complex left and right
coordinates $X_{\mathrm{L}}=ia(1,0,0,1)$ and $X_{\mathrm{R}}=ia(1,0,0,-1).$
 The time-like components of the both sources are adjoined to the same
right end of the complex string interval $[-ia,ia].$ This
asymmetry is removed by a remarkable stringy construction -
orientifold \cite{Hor,BurStr,BurOri,BurTwi}.

\subsection{Orientifold}

The models of relativistic strings contain usually two stringy
modes: left and right. So, the modes
$X_{\mathrm{L}}(\t_{\mathrm{L}})$ and
$X_{\mathrm{R}}(\t_{\mathrm{R}})$ represent only the half-strings
on the interval $\sigma \in [-a,a].$ Orientifold is formed from
two open half-strings which are joined forming one closed string,
and this closed string is to be folded one. The interval $[-a,a]$
is covered by parameter $\sigma$ twice:  the first time from left
to right, and (say) the left half-string has the usual
parametrization, while the interval $[-a,a]$ is reversed and
covers the original one in opposite orientation for the right
half-string. Therefore, the string is formed by two half-strings
and turns out to be closed, but folded. The right and left string
modes  are flipping between the the initiate and the reversed
intervals. One sees that for the complex interval the revers is
equivalent to complex conjugation of the parameter $\t. $  So, on
e
has to put $\t_{\mathrm{R}} = \bar \t _{\mathrm{L}} .$\fn{Details
of this construction may be found in
\cite{Hor,BurStr,BurOri,BurTwi}.} After orientifolding, the
complex timelike coordinates of the points  $X_{\mathrm{L}}$ and
$\bar X_{\mathrm{R}}$ turns out to be sitting on the opposite ends
of the interval $[-a,a],$ while their imaginary space-like
coordinates will be coinciding, which corresponds to one of the
necessary orientifold condition $X_{\mathrm{L}}(\t_{\mathrm{L}}) =
\bar X_{\mathrm{R}} (\bar \t _{\mathrm{R}})$.

\section{Conclusion}

The above treatment shows that there is a deep internal
relationships between the Dirac equation and the complex
representation of the Kerr geometry. The Dirac equation works in
the complex Minkowski space-time, and electron is not elementary
point-like object, but has a many-sided structure. It has a
nontrivial complex structure which is related to the real and
complex structures of the Kerr geometry. The space-time source of
the naked electron represents a very specific complex string with
two point-like (quark-like) sources located on the ends of this
string.  In the same time, after orientifolding this string, the
space coordinates of these sources are merging, turning into {\it
one complex point} shifted in the imaginary direction on the
Compton distance $a$.  This complex position of the source is,
apparently, the origin of the problems with the position operator
in the Dirac theory, along with the complicate topology of the
real Kerr geometry, and presence of the string-like excitations.

The obtained recently multiparticle  Kerr-Schild solutions
\cite{Multi} shed some light on the multiparticle structure of the
 considered in QED dressed electron. This treatment is based on the
remarkable properties of the Kerr theorem. There is also
remarkable renormalization of the Kerr singularity by
gravitational field \cite{renorm}. However, these questions go out
of the frame of this paper.

 \section*{Acknowledgments}

We are thankful to  V. Arcos, A. Khrennikov, J. Pe
reira, I.
Volovich, and especially to Prof. G. 't Hooft for very
illuminating discussions. We are also thankful
to T. Nieuvenhuisen for the attention to this work and
stimulating discussions on the Dirac theory.

\end{document}